\documentclass[11pt]{article}
\usepackage[utf8]{inputenc}
\usepackage{graphicx}
\usepackage{graphics,epstopdf}
\usepackage{amssymb,amsmath}
\usepackage{booktabs} 
\usepackage{xcolor}
\usepackage{cite}
\usepackage{caption}
\usepackage{float}
\usepackage{booktabs}
\usepackage{subcaption}
\usepackage{soul}       % For [H] placement

\usepackage{geometry} % For adjusting page margins

\usepackage[a4paper=true,pagebackref=false]{hyperref} %true
\hypersetup{colorlinks = true, linkcolor = red, anchorcolor = red, citecolor = blue, filecolor = red, pagecolor = red, urlcolor = red}

\title{Late-Time Cosmic Acceleration in Ho\v{r}ava-Lifshitz Gravity: 
Observational Evidence from Cosmic Chronometers and Pantheon+SHOES Datasets}

\author{ Shivani Sharma\footnote{\textbf{corresponding author}: Shivani Sharma, shivani1506@bhu.ac.in}, R. Chaubey\footnote{yahoo\underline{ }raghav@rediffmail.com, rchaubey@bhu.ac.in} \\
Centre for Interdisciplinary Mathematical Sciences \\Institute of Science, Banaras Hindu University\\ Varanasi, Pin 221005, India}

\begin{document}

\maketitle

\begin{abstract}
We investigate the cosmological implications of Horava-Lifshitz (HL) gravity using a redshift-dependent deceleration parameter of the form
    $q(z) = q_0 + \frac{q_1 \ln(1+z)}{1 + n\ln(1+z)}$, from which the Hubble parameter $H(z)$ is derived analytically. This parametrization captures the transition from early deceleration to late-time acceleration, with a logarithmic correction governed by $n$ that distinguishes it from standard kinematic models. Model parameters $H_0$, $q_0$, $q_1$, and $n$ are constrained via MCMC using cosmic chronometer (CC) and Pantheon+SHOES datasets, individually and in combination. Across all dataset combinations, $q_0 < 0$, confirming ongoing accelerated expansion. The reconstructed $H(z)$ is consistent with $\Lambda$CDM at low redshifts, with mild deviations at higher redshifts. The $\{r, s\}$ statefinder parameters indicate that the model evolves smoothly, beginning with Chaplygin gas-type behaviour, crossing the $\Lambda$CDM fixed point, and settling into a quintessence-like phase at late times. The $Om(z)$ diagnostic reveals negative slopes throughout, 
indicating quintessence-like dark energy ($w > -1$).  Present-day cosmic age estimates from the individual and combined datasets yield $t_0 \approx 13.7$ Gyr, in agreement with Planck 2018 constraints.
\end{abstract} 

\section{Introduction}
One of the most fascinating open questions in modern cosmology is the observed accelerated expansion of the universe at late times. This phenomenon was first confirmed through luminosity distance observations of Type Ia supernovae (SNe Ia) \cite{Riess1998, Perlmutter1999} and has since been supported by various independent probes, including cosmic microwave background (CMB) \cite{Planck2020} anisotropies, baryon acoustic oscillations (BAO) \cite{Eisenstein2005}, and large-scale structure surveys \cite{Tegmark2004}. In the framework of standard cosmology, the accelerated expansion is attributed to the cosmological constant $\Lambda$, representing vacuum energy density. Despite its strong observational support, the $\Lambda$CDM model faces two major theoretical concerns: the fine-tuning problem, arising from the large gap between theoretical predictions and observations, and the coincidence problem, questioning why matter and dark energy densities are comparable today \cite{Weinberg1989, Sahni2000}. These unresolved issues have encouraged researchers to explore a wide range of alternative frameworks that account for cosmic acceleration beyond the standard cosmological constant.
\par A promising class of alternatives is given by modified theories of gravity, where cosmic acceleration arises from geometric modifications to the Einstein–Hilbert action rather than from an exotic form of dark energy. A variety of modified gravity theories have been proposed and extensively analyzed, such as $f(R)$ \cite{Nojiri2011}, $f(T)$ \cite{Ferraro2007}, and $f(Q)$ \cite{Jimenez2018} gravity, together with scalar–tensor approaches like Brans–Dicke theory \cite{Brans1961}, each reflecting different modifications of general relativity (GR). Within this class of theories, Ho\v{r}ava–Lifshitz (HL) gravity, proposed by Ho\v{r}ava, stands out as a compelling and widely studied model \cite{Horava2009}. HL gravity introduces an anisotropic scaling of space and time at high energy scales, which ensures power-counting renormalizability and helps to address the problem of quantum gravity, while reducing to GR in the infrared regime \cite{Sotiriou2009}. Numerous studies have examined the cosmological behavior of HL gravity, revealing its potential to drive late-time acceleration, generate bouncing cosmologies, and resolve the horizon problem without relying on inflation \cite{Wei2011, Calcagni2009, Kiritsis2009, Brandenberger2009, Lu2009, 
Mukohyama2009}. Despite their advantages, many HL cosmological models face a notable limitation: they require specific fluid or scalar field assumptions to close the system of equations, leading to additional free parameters with limited observational justification \cite{Saridakis2010, 
Chaudhary2024, Paul2013, Biswas2015}. The present study is motivated by the need to address and overcome these limitations.
\par To address this, instead of assuming any specific matter content or scalar field setup, we follow a model-independent kinematic approach by directly parametrizing the deceleration parameter $q(z)$. This allows us to reconstruct the expansion history purely from observations, without introducing extra dynamical assumptions. In recent years, kinematic reconstructions have attracted significant attention. One of the simplest forms is the linear parametrization  $q(z) = q_0 + q_1 z $, however, it diverges at high redshift, which restricts its range of applicability \cite{Cunha2008}. To overcome this issue, a divergence-free form $q(z) = q_0 + q_1 z/(1+z)$ \cite{Xu2008} was proposed, inspired by the CPL parametrization \cite{Chevallier2001, Linder2003}. Although it resolves the divergence problem, it still lacks enough flexibility to describe finer details of the transition epoch. A logarithmic parametrization of $q(z)$   was proposed by Mamon and Das \cite{Mamon2017} to investigate the cosmic expansion using SNIa, BAO, and CMB data. More recently, Chaudhary et al.\cite{Chaudhary2023} systematically compared different  $q(z)$ models using MCMC methods with CC, BAO, and Pantheon$^{+}$ datasets. Furthermore, Bhoyar and Ingole \cite{Bhoyar2024} studied a logarithmic form in $f(Q,C)$  gravity, whereas Samaddar and Singh \cite{Samaddar2025} also considered a logarithmic form and study the  behavior in  Ho\v{r}ava-Lifshitz  gravity. However, these parametrizations have not yet been systematically examined within the framework of HL gravity. In this work, we consider a logarithmic form given by $q(z) = \frac{q_0 + q_1 \ln(1+z)}{[1 + n\ln(1+z)]}$, which remains bounded at all redshifts and features a parameter $n$ that controls the saturation of the logarithmic contribution. In this parametrization, the evolution naturally interpolates between the present value $q_0$ and an early-time asymptotic behavior governed by $q_1/n$, resulting in a more expressive and analytically convenient description of the transition era. In this work, we employ this flexible kinematic ansatz within the framework of HL gravity to investigate late-time cosmological behavior, keeping the analysis largely independent of specific fluid or field models, while being strongly constrained by observational datasets, namely cosmic chronometers (CC), Pantheon+SHOES, and their joint analysis.
\par 
The paper is organized in the following way. Section \ref{s2} presents the basic mathematical framework of Hořava–Lifshitz gravity, including   the form of the gravitational action under detailed balance and the modified Friedmann equations. In this section, we also propose a logarithmic form for the deceleration parameter and obtain the related Hubble parameter $H(z)$. In Section \ref{s3}, we discuss the observational data, including cosmic chronometers (CC) and the Pantheon+SH0ES compilation, along with the MCMC approach used to estimate the model parameters. In Section \ref{s4}, we explore the behaviour of various cosmological quantities, including energy density, pressure, the deceleration parameter, and the equation of state. We further examine the statefinder $\{r, s\}$ diagnostic, the $Om(z)$ diagnostic, and the age of the Universe. In the end, Section \ref{s5} summarizes the key findings of this work and provides the overall conclusions.

\section{Mathematical Framework of Ho\v{r}ava–Lifshitz Gravity}\label{s2}
In our formulation, we consider the Arnowitt–Deser–Misner decomposition of the metric, which takes the form \cite{Calcagni2009,calcagni2010, kiritsis2010}
\begin{equation}\label{1}
    ds^2 = -N^2 dt^2 + g_{ij}(dx^i + N^i dt)(dx^j + N^j dt)
\end{equation}
In this framework, $N$ represents the lapse function, $N_i$ denotes the shift vector, and $g_{ij}$ defines the metric tensor. The scaling behavior is given by $t \to l^3 t$ and $x^i \to l x^i$. The total action of HL gravity consists of kinetic and potential terms, each with its own physical role.
\begin{equation}\label{2}
    S_g = S_k + S_v = \int d^3x \, dt \sqrt{g} \, N (\mathcal{L}_k + \mathcal{L}_v)
\end{equation}
The kinetic term can be evaluated as
\begin{equation} \label{3}
    S_k = \int d^3x \, dt \sqrt{g} \, N \left[ \frac{2(K_{ij}K^{ij} - \lambda K^2)}{\kappa^2} \right]
\end{equation}
where the extrinsic curvature is defined as
\begin{equation} \label{4}
    K_{ij} = \frac{\dot{g}_{ij} - \Delta_i N_j - \Delta_j N_i}{2N}
\end{equation}
The symmetry of the Lagrangian $\mathcal{L}$ helps in reducing the number of invariants \cite{hovrava2009, hovrava2009a, lifshitz}. This symmetry is called detailed balance, and by applying it, the action can be expressed in an expanded form as follows.
\begin{equation} \label{5}
    \begin{split}
S_g = \int d^3x \, dt \sqrt{g} \, N \Bigg[ & \frac{2(K_{ij}K^{ij} - \lambda K^2)}{\kappa^2} 
+ \frac{\kappa^2 C_{ij} C^{ij}}{2\omega^4} 
- \frac{\kappa^2 \mu \epsilon^{ijk} R_{i,j} \Delta_j R^m_{\ k}}{2\omega^2 \sqrt{g}} \\
& + \frac{\kappa^2 \mu^2 R_{ij} R^{ij}}{8} 
- \frac{\kappa^2 \mu^2}{8(3\lambda - 1)} 
\left( \frac{(1 - 4\lambda)R^2}{4} + \Lambda R - 3\Lambda^2 \right) \Bigg]
\end{split}
\end{equation}

where the cotton tensor $C^{ij}$ is given by
\begin{equation} \label{6}
    C^{ij} = \frac{\epsilon^{ijk} \nabla_k \left( R^j_{\ i} - \frac{1}{4} R \delta^j_i \right)}{\sqrt{g}}
\end{equation}
All covariant derivatives are defined with respect to the spatial metric $g_{ij}$, and $\epsilon^{ijk}$ denotes the totally antisymmetric unit tensor. Here, $\lambda$ is a dimensionless constant, while $\kappa$, $\omega$, and $\mu$ are fixed parameters. By considering the lapse function to be  dependent only on time, $N \equiv N(t)$, Ho\v{r}ava derived the corresponding gravitational action. Adopting the FLRW metric with $N = 1$, $g_{ij} = a^2(t)\gamma_{ij}$, $N_i = 0$, and $\gamma_{ij} dx^i dx^j = \frac{dr^2}{1 - kr^2} + r^2 d\Omega^2$, the cosmological framework is established.\\
Here, the values $k = -1, 1, 0$ describe open, closed, and flat Universes, respectively. The Friedmann equations \cite{jamil2010, paul2012} are derived by varying the action with respect to $N$ and $g_{ij}$.
\begin{equation}\label{7}
    H^2 = \frac{\kappa^2 \rho}{6(3\lambda - 1)} + \frac{\kappa^2}{6(3\lambda - 1)} 
\left[ \frac{3\kappa^2 \mu^2 \Lambda^2}{8(3\lambda - 1)} 
+ \frac{3\kappa^2 \mu^2 k^2}{8(3\lambda - 1) a^4} \right] 
- \frac{\kappa^4 \Lambda \mu^2 k}{8(3\lambda - 1)^2 a^2}
\end{equation}

\begin{equation} \label{8}
    2\dot{H} + 3H^2 = -\frac{\kappa^2 p}{2(3\lambda - 1)} 
- \frac{\kappa^2}{2(3\lambda - 1)} 
\left[ \frac{3\kappa^2 \mu^2 \Lambda^2}{8(3\lambda - 1)} 
+ \frac{3\kappa^2 \mu^2 k^2}{8(3\lambda - 1) a^4} \right] 
- \frac{\kappa^4 \Lambda \mu^2 k}{4(3\lambda - 1)^2 a^2}
\end{equation}

A notable feature of HL gravity is the presence of a term proportional to $1/a^4$, which can be treated as a ``dark radiation" term \cite{calcagni2010, kiritsis2010}, while the constant term plays the role of the cosmological constant. The Hubble parameter is given by $H = \dot{a}/a$, where the dot denotes differentiation with respect to cosmic time $t$.\\
We define the deceleration parameter through the relation:
\begin{equation}\label{9}
    q = -1 - \frac{\dot{H}}{H^2}
\end{equation}
The expansion history of the Universe is governed by the sign of $q$, where $q > 0$ implies a decelerating phase and $q < 0$ an accelerating one. By solving Eq.$(\ref{9})$, we get:
\begin{equation} \label{10}
    H(z) = H_0 \exp\left[\int_0^z \frac{1 + q(z')}{1 + z'} \, dz'\right]
\end{equation}
By making the suitable choice of  the functional form of $q(z)$, one can effectively probe the evolution of key cosmological quantities such as the Hubble parameter and the equation of state parameter. This not only helps in understanding their time dependence but also provides meaningful insights into the expansion history of the Universe. Based on this idea, we propose
\begin{equation} \label{11}
    q(z) = q_0 + \frac{q_1 \ln(1+z)}{1 + n \ln(1+z)}
\end{equation}

Here, $q_0$ represents the present-day value of the deceleration parameter, while $q_1$ determines how $q(z)$ evolves with redshift. The parameter $n$ is a dimensionless constant that controls the saturation behavior of the logarithmic term. The choice of this form is driven by the need to describe the universe's expansion in a smooth and adaptable way, without encountering divergences across the full range of redshift. The adopted functional form reflects important theoretical considerations and observational relevance, making it well-suited for probing the cosmological dynamics within HL gravity. The sign of $q_1$ plays a crucial role: a positive $q_1$ enables a transition from deceleration to acceleration, while a negative $q_1$ results in a monotonic behavior. From a phenomenological perspective, this form of $q(z)$ smoothly connects two important regimes. At low redshift ($z \ll 1$), using $\ln(1+z) \approx z$, the expression simplifies to $q(z) \approx q_0 + q_1 z/(1 + nz) \approx q_0 + q_1 z$. This behavior closely resembles the commonly used linear parametrizations and allows for direct comparison with established models such as CPL (Chevallier-Polarski-Linder). Moreover, for large redshift values ($z \gg 1$), the ratio $\ln(1+z)/[1 + n\ln(1+z)]$  asymptotically approaches the finite value $1/n$, implying that $q(z)$ approaches $q_0 + q_1/n$. This guarantees that the deceleration parameter stays bounded and aligns with the expected matter-dominated decelerating phase of the early Universe and avoids any unphysical divergence or oscillations. The value of $q_0 + q_1/n$ at high redshift can be tuned to match $q \approx 1/2$, which is consistent with the matter-dominated epoch and necessary for the formation of large-scale structures. Secondly, the proposed form makes it straightforward to determine the transition redshift $z_t$, defined through $q(z_t) = 0$, indicating the change from deceleration to acceleration. Solving the relation $q_0 + q_1 \ln(1+z_t)/[1 + n\ln(1+z_t)] = 0$ leads to the closed-form solution $z_t = \exp[-q_0/(q_1 + nq_0)] - 1$. This provides a simple way to locate the transition epoch, which observationally lies around $z_t \approx 0.6$-$0.8$. The simplicity and analytical clarity of the transition behavior make this parametrization particularly appealing. In addition, the parameter $n$ governs the rate at which the logarithmic correction saturates, with smaller $n$ allowing a slower growth and larger $n$ leading to a rapid saturation. This flexibility allows the three-parameter model $(q_0, q_1, n)$ to go beyond the limitations of standard two-parameter forms, allowing it to describe finer details of the transition from deceleration to acceleration.
\par It should be noted that this formulation is not directly obtained from the fundamental Ho\v{r}ava–Lifshitz field equations. Even so, it should not be viewed as purely ad hoc. Rather, it is an observationally motivated and phenomenologically sound ansatz that provides a useful way to reconstruct the expansion history in the context of the modified Friedmann equations of HL gravity. A logarithmic dependence on redshift fits well within this context, since Ho\v{r}ava–Lifshitz gravity is built upon anisotropic scaling principles. This key feature, where space and time scale differently at high energies, naturally produces scale-dependent modifications to gravitational dynamics. Constructing the deceleration parameter using $\ln(1+z)$ ensures a smoother evolution compared to polynomial approaches. This feature helps in capturing mild, scale-dependent deviations from general relativity, while keeping the infrared regime intact and consistent with observations. Another key point is that the parametrization is purely kinematic in nature. This means that no prior assumptions about matter components, scalar field potentials, or equations of state are required, making it a flexible and theory-independent tool that can be implemented in HL gravity and other models. In this way, the model offers a clean framework to compare HL cosmology with current precision datasets, including cosmic chronometers, and Pantheon$+$ SHOES. At the same time, it remains free from the additional assumptions and poorly constrained parameters that usually arise in dynamical dark energy scenarios.\\
By making use of Eqs. (\ref{10}) and (\ref{11}), one can derive
\begin{equation}\label{12}
    H(z) = H_0 (1+z)^{(1+q_0)} \Bigg[\frac{(1+z)^{q_1/n}}{(1+n\ln(1+z))^{q_1/n^2}}\Bigg]
\end{equation}
Next, we compare the predictions of our model with observational data from different cosmological probes, including Hubble parameter measurements and SNe Ia data from the Pantheon+SHOES compilation, to constrain the parameters $H_0$, $q_0$, $q_1$, and $n$.

\par This work brings a novel perspective to HL gravity by combining its modified Friedmann equations with observational data via a new deceleration parameter, $q(z) = q_0 + q_1 \ln(1+z)/[1 + n\ln(1+z)]$. It provides a model-independent way to reconstruct the expansion history without introducing dark energy assumptions. While many previous approaches assumed a fixed form of $H(z)$ or specific matter components, our framework avoids such choices by relying on a kinematic ansatz. Its logarithmic form and saturation mechanism provide a natural description of the scale-dependent features of HL gravity. The three-parameter structure of the model makes it possible to independently describe both the present accelerated expansion and the earlier decelerating phase, which is not achievable with simpler two-parameter forms. The strong consistency of the constrained parameters with cosmic chronometer and Pantheon$^+$ data, both individually and jointly, shows that the infrared limit of HL gravity can effectively explain the observed late-time acceleration, supporting it as a viable alternative to $\Lambda$CDM.

\section{Data Sets and Methodological Framework}\label{s3}
To validate the model against observational evidence, we consider the following datasets:
\subsection{The cosmic chronometer data}
The cosmic chronometer (CC) dataset provides direct measurements of the expansion rate of the Universe. This method describes the expansion of the Universe using the scale factor $a$ and its relation to redshift $z$. The Hubble parameter is then defined as $H(z) = -\frac{1}{1+z}\frac{dz}{dt}$ \cite{jimenez2002}. The CC dataset used here contains 31 measurements of the Universe’s expansion rate across the redshift range $0.07 \leq z \leq 1.965$ \cite{vagnozzi2021}. To estimate the likelihood from the CC dataset, we define the $\chi^2$ function as \cite{singh2024affine}
 \begin{equation}\label{13}
\chi_c^2 = \sum_{i=1}^{N} \frac{\left[ H_t(z_i, \Theta_p) - H_o(z_i) \right]^2}{\sigma^2(z_i)} .
\end{equation}

The terms $H_t(z_i, \Theta_p)$ and $H_o(z_i)$ denote the theoretical and measured values of the Hubble parameter at redshift $z_i$, respectively, whereas $\sigma(z_i)$ gives the uncertainty in the observed data.

\subsection{Pantheon + SHOES data}
In this work, we use Type Ia supernova (SNIa) data from the Pantheon+SH0ES compilation, which contains 1701 data points covering the redshift range $0.00122 \leq z \leq 2.26137$. This dataset includes $1701$ light curves corresponding to $1550$ spectroscopically confirmed SNe Ia collected from 18 different surveys \cite{brout2022, scolnic2022}. For our analysis, we take the apparent magnitude $m_b^{\text{corr}}$ from the `m\_b\_corr' column of the Pantheon+SH0ES data. We follow the methodology described in Brout et al. \cite{brout2022} for our analysis. The distance modulus $\mu(z)$ is connected to the luminosity distance $d_l$ via $\mu(z) = 5 \log_{10}(d_l(z)/\text{Mpc}) + 25$, where $d_l$ is computed as $d_l = c(1 + z_{\text{hel}})\int_{0}^{z_{\text{HD}}} \frac{dz}{H(z)}$. The parameters $z_{\text{HD}}$ and $z_{\text{hel}}$ are extracted from the ‘zHD’ and ‘zhel’ columns of the Pantheon+SH0ES dataset, respectively. The parameter $z_{\text{hel}}$ indicates the redshift measured in the heliocentric frame, while $z_{\text{HD}}$ denotes the corrected redshift that accounts for peculiar galaxy motions and represents the cosmic rest frame value \cite{brout2022, scolnic2022}. Parameter estimation for the model is carried out by minimizing the $\chi^2$ function, as defined in \cite{brout2022, scolnic2022, singha2025, singh2026}.
\begin{equation}
\chi^2_{ps} = \Delta Q^T C^{-1} \Delta Q, 
\label{15}
\end{equation}
In this expression, $C^{-1}$ represents the inverse covariance matrix with dimensions $1701 \times 1701$, and the elements of $Q$ are defined as in \cite{brout2022, scolnic2022}.
\begin{equation}
Q_i =
\begin{cases}
m_i - M - \mu^{\mathrm{cep}}_i, & \text{if } i \in \text{Cepheid hosts}, \\
m_i - M - \mu_{\mathrm{th}}(z_i), & \text{otherwise}.
\end{cases}
\label{16}
\end{equation}

Here, $\mu^{\mathrm{cep}}_i$ represents the distance modulus of the Cepheid host and is associated with the ‘IS\_CALIBRATOR’ column in the Pantheon+SH0ES data.
\par
\begin{figure}[h!]
\centering
		\includegraphics[height=7.0cm,width=10.5cm]{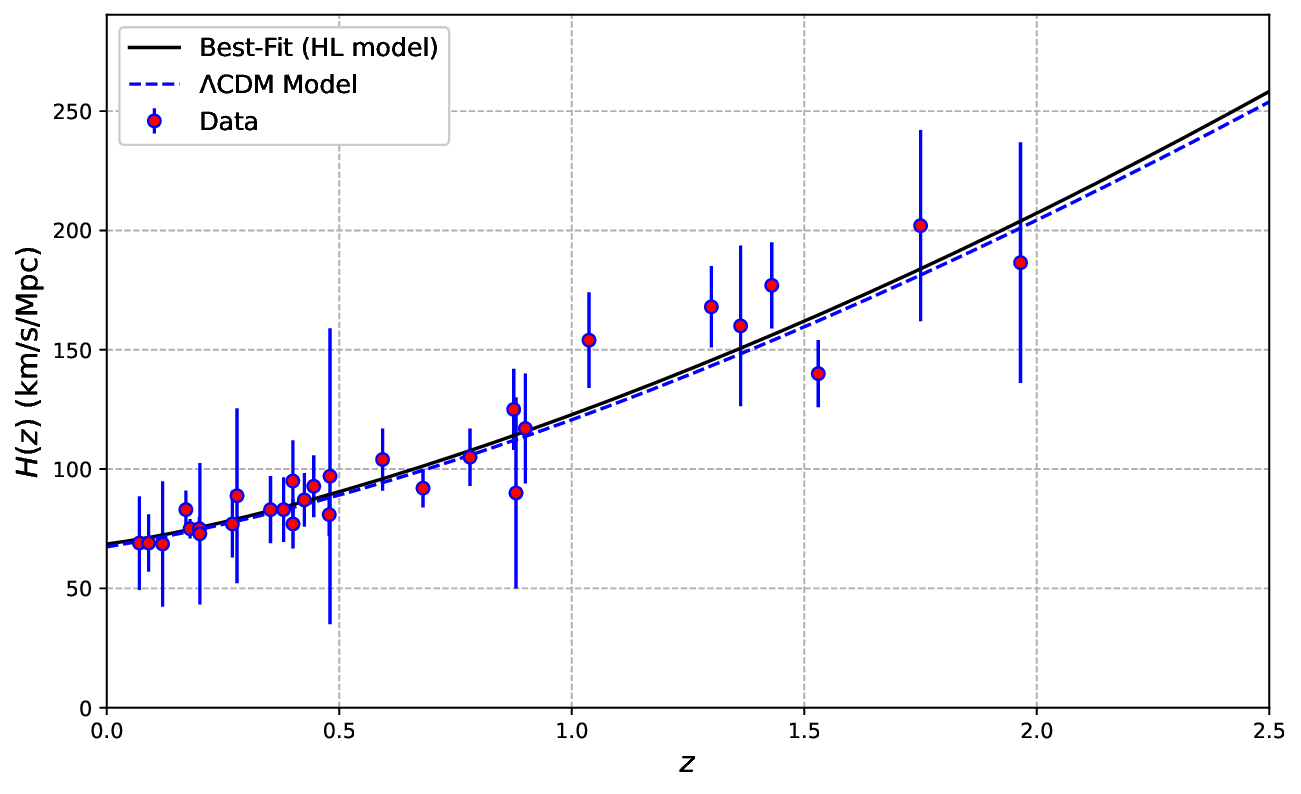}
		\caption{The best-fit $H(z)$ curve with cosmic chronometer data and its error bars. } 
\label{f1}
\end{figure}

\begin{figure}[h!]
\centering
		\includegraphics[height=7.5cm,width=10.5cm]{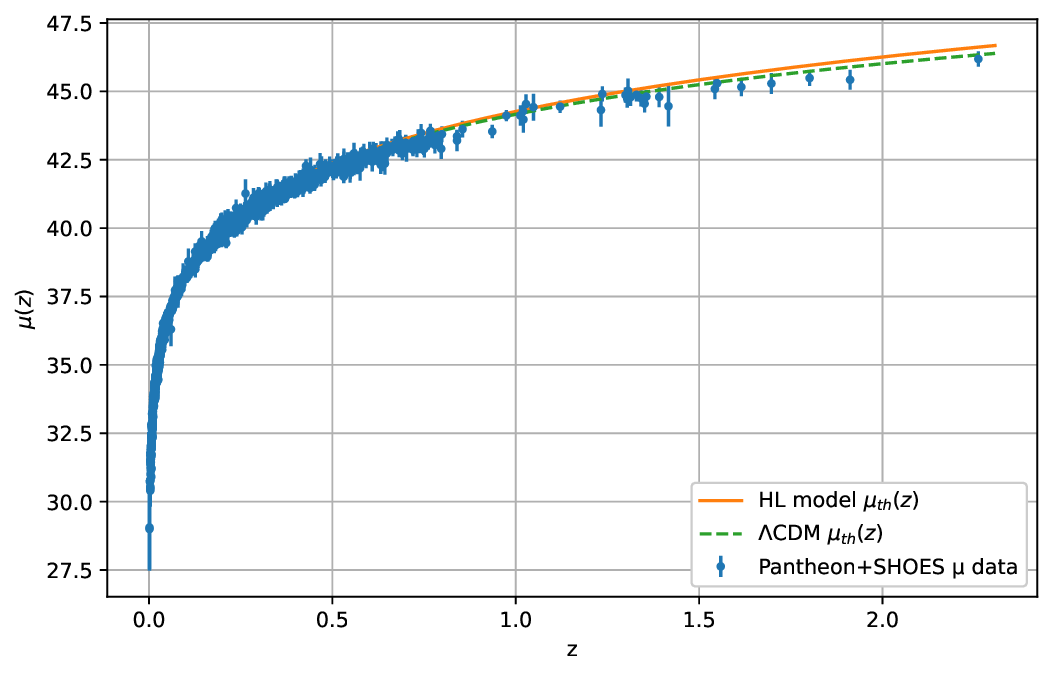}
		\caption{The best-fit $H(z)$ curve with Pantheon+ SHOES data and its error bars.} 
\label{f2}
\end{figure}

\begin{figure}[h!]
\centering
		\includegraphics[height=10.0cm,width=8.5cm]{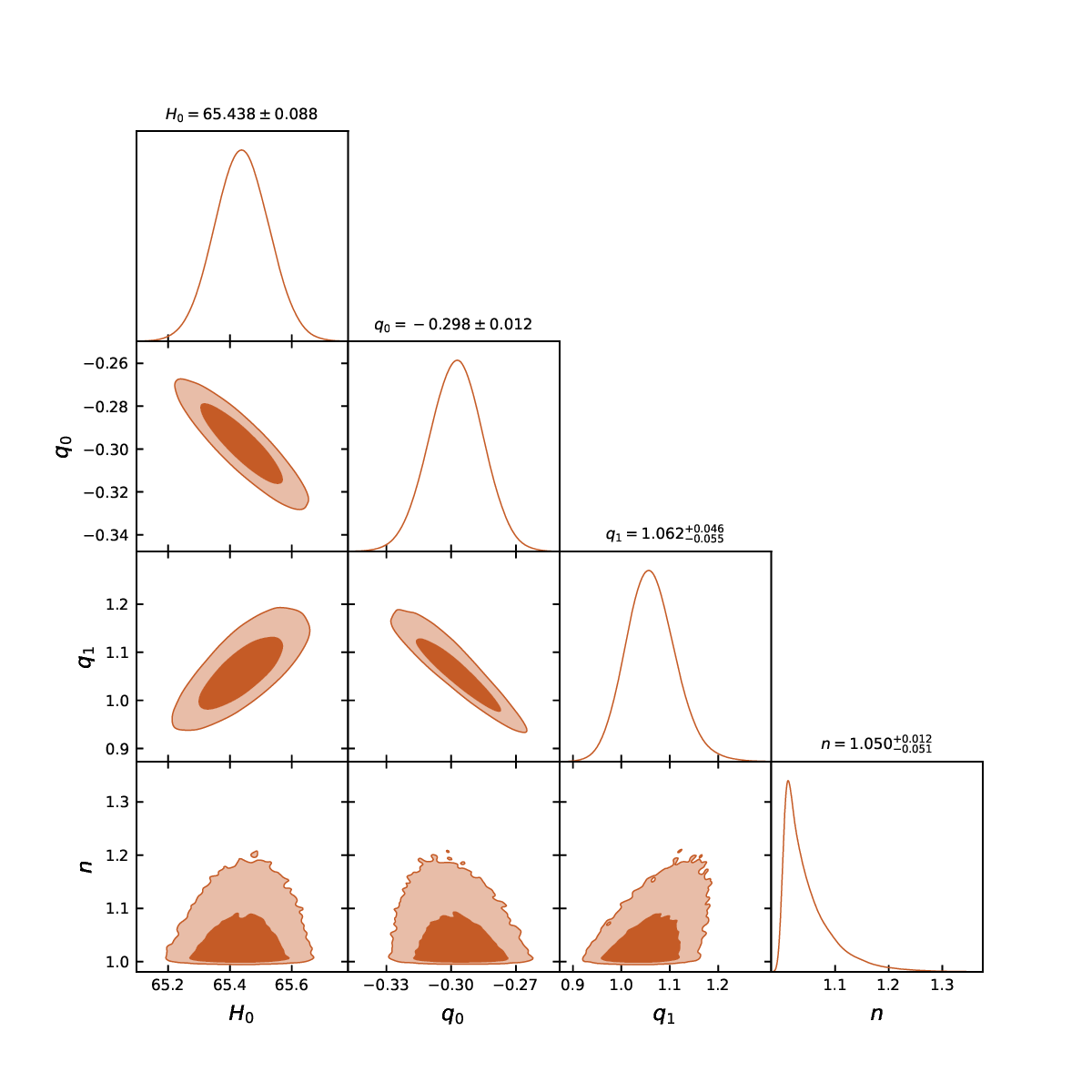}
		\caption{The posterior distribution in the form of 2D contour and 1D plot for HL model using the cosmic chronometer data. } 
\label{f3}
\end{figure}

\begin{figure}[h!]
\centering
		\includegraphics[height=10.0cm,width=10.0cm]{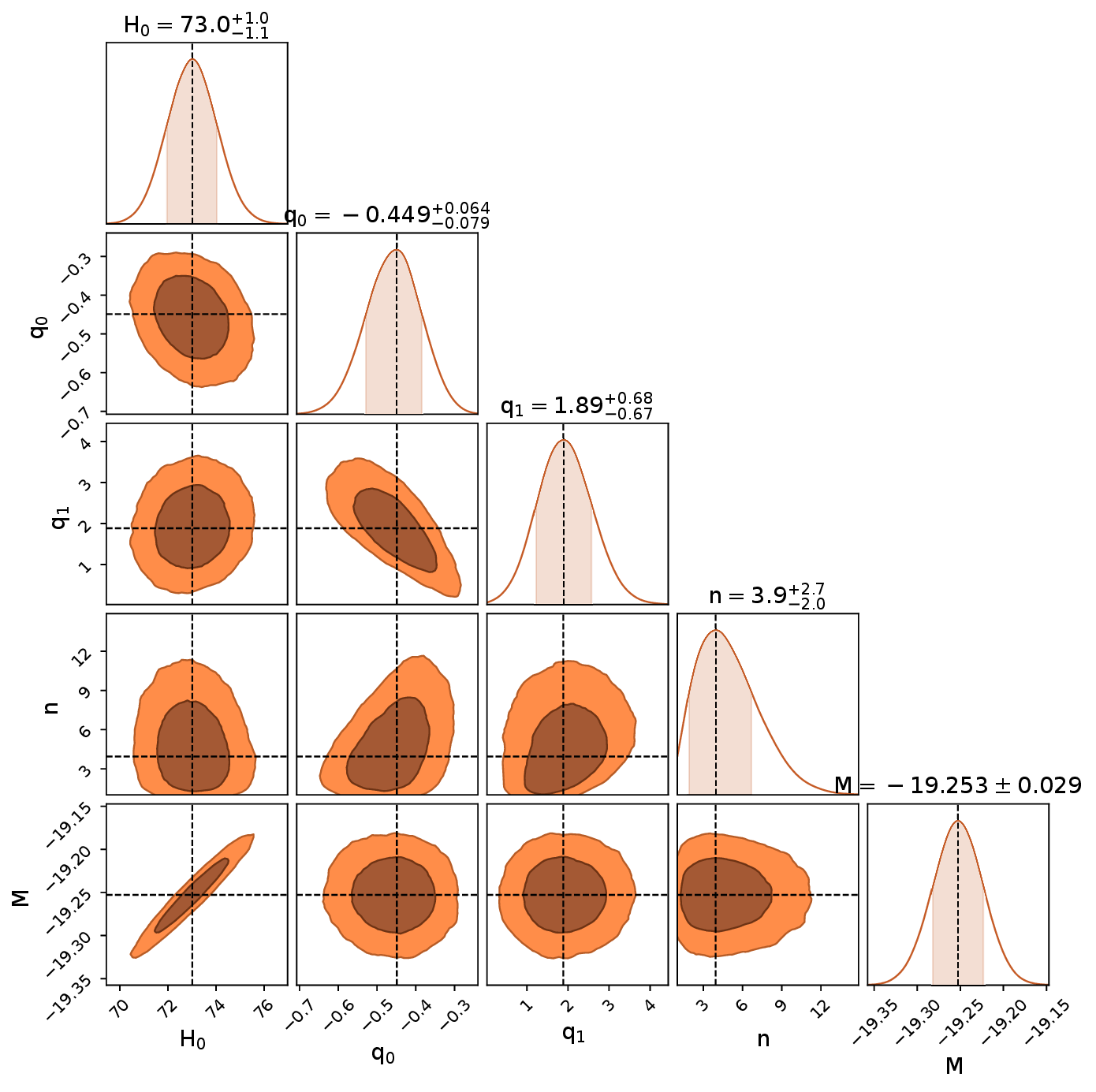}
		\caption{The posterior distribution in the form of 2D contour and 1D plot for HL model using the Pantheon+SHOES data. } 
\label{f4}
\end{figure}

\begin{figure}[h!]
\centering
		\includegraphics[height=10.0cm,width=10.0cm]{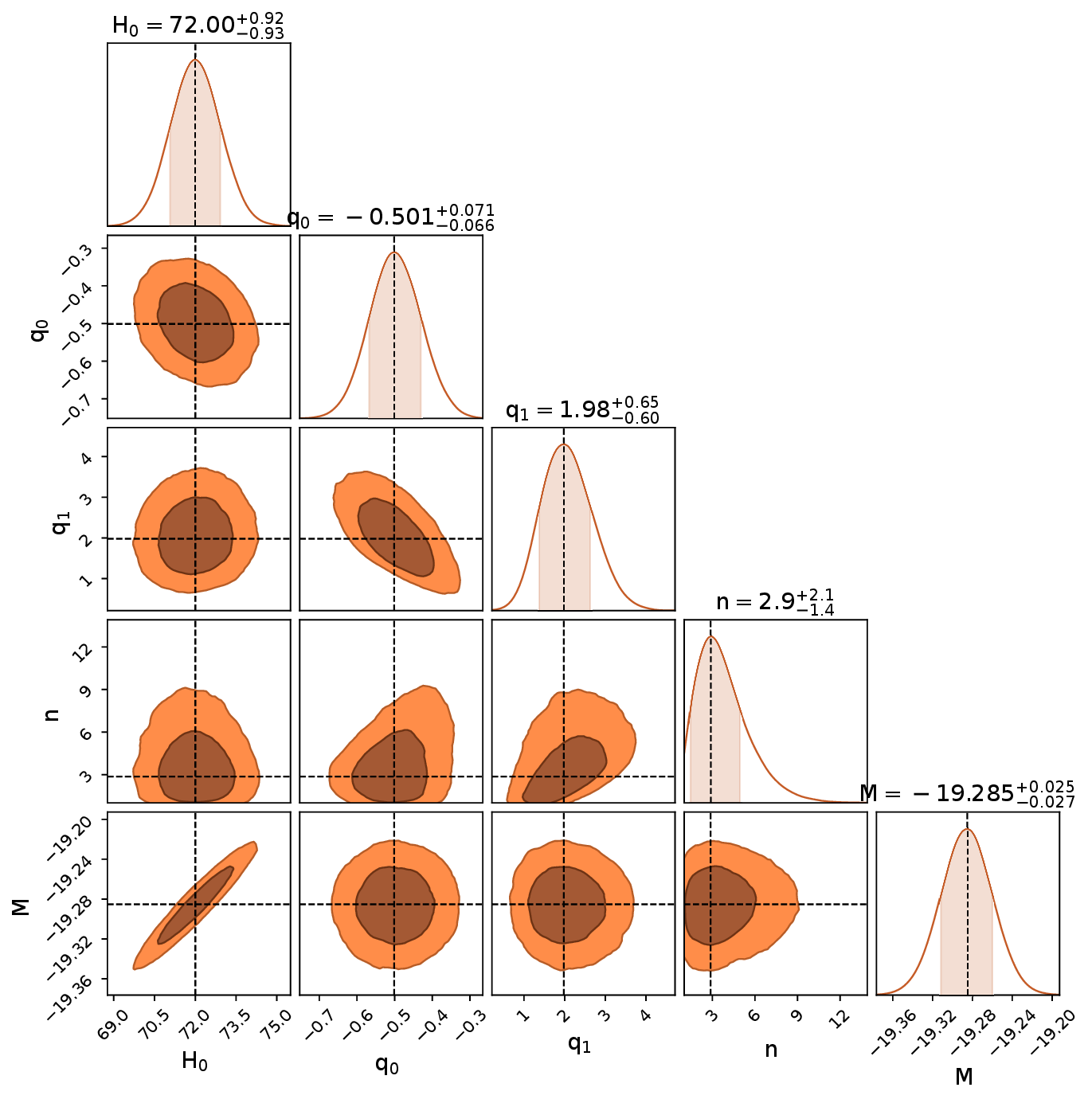}
		\caption{The posterior distribution in the form of 2D contour and 1D plot for HL model using the CC+Pantheon+SHOES data. } 
\label{f5}
\end{figure}

 In this work, we estimate the model parameters using CC and Pantheon+SH0ES datasets, both individually and in combination (CC + Pantheon+SH0ES). We employ Bayesian analysis based on the Markov Chain Monte Carlo (MCMC) technique to obtain robust constraints. The sampling of the parameter space is performed using the \texttt{emcee} package \cite{foreman2013emcee}. The combined chi-square function is written as
\begin{equation}
    \chi^2_t = \chi^2_{cc} + \chi^2_{ps} 
\end{equation}
The phase planes constructed from the 1D marginalized posteriors and 2D contour plots of the model parameters are presented in Figure \ref{f3} for the cosmic chronometer dataset. For the CC analysis, the constrained values are $H_0 = 65.438 \pm 0.088$, $q_0 = -0.298 \pm 0.012$, $q_1 = 1.062^{+0.181}_{-0.055}$, and $n = 1.050^{+0.126}_{-0.063}$. Similarly, Figure \ref{f4} presents the posterior distribution for the Pantheon$^{+}$SH0ES dataset, yielding the best-fit values $H_0 = 73.0^{+1.0}_{-1.1}$, $q_0 = -0.449^{+0.064}_{-0.079}$, $q_1 = 1.89^{+0.68}_{-0.67}$, $n = 3.9^{+2.7}_{-2.0}$, and $M = -19.253 \pm 0.029$. The joint analysis using CC + Pantheon$^{+}$SH0ES data is shown in Figure \ref{f5}, which gives $H_0 = 72.00^{+0.92}_{-0.93}$, $q_0 = -0.501^{+0.071}_{-0.066}$, $q_1 = 1.98^{+0.65}_{-0.60}$, $n = 2.9^{+2.1}_{-1.4}$, and $M = -19.285^{+0.025}_{-0.027}$. In all three cases, the negative values of $q_0$ firmly establish the present-day accelerated expansion of the Universe.  For the analysis of all three datasets, we impose uniform priors defined by
$40 < H_0 < 90$, $-2 < q_0 < 2$, $1.001 < n < 15$, and $-20 < M < -18$. A consolidated summary of the best-fit parameter values obtained from each dataset is provided in Table~\ref{t1}. Figure \ref{f1} displays the reconstructed Hubble parameter $H(z)$ as a function of redshift for the CC dataset, where the best-fit curve of the HL model is compared against the standard $\Lambda$CDM prediction and the observational data points with associated error bars. A strong agreement between the two models is observed at low redshifts, with slight deviations emerging at higher redshifts. Figures \ref{f2} and present the corresponding comparison for the Pantheon$^{+}$SH0ES dataset, showing the theoretical distance modulus $\mu(z)$ of the HL model alongside the $\Lambda$CDM curve and the observed supernova data. The close overlap between the model prediction and the data across the entire redshift range confirms the observational viability of the proposed parametrization within the Ho\v{r}ava-Lifshitz gravity framework.

\begin{table}[h!]
\centering
\caption{Summary of the best-fit parameter values with $1\sigma$ uncertainties obtained from the CC, Pantheon$^{+}$SH0ES, and joint CC + Pantheon$^{+}$SH0ES datasets.}
\label{t1}
\begin{tabular}{lccc}
\hline\hline
Parameter & CC & Pantheon$^{+}$SH0ES & CC + Pantheon$^{+}$SH0ES \\
\hline
$H_0$ & $65.438 \pm 0.088$ & $73.0^{+1.0}_{-1.1}$ & $72.00^{+0.92}_{-0.93}$ \\[6pt]
$q_0$ & $-0.298 \pm 0.012$ & $-0.449^{+0.064}_{-0.079}$ & $-0.501^{+0.071}_{-0.066}$ \\[6pt]
$q_1$ & $1.062^{+0.181}_{-0.055}$ & $1.89^{+0.68}_{-0.67}$ & $1.98^{+0.65}_{-0.60}$ \\[6pt]
$n$ & $1.050^{+0.126}_{-0.063}$ & $3.9^{+2.7}_{-2.0}$ & $2.9^{+2.1}_{-1.4}$ \\[6pt]
$M$ & -- & $-19.253 \pm 0.029$ & $-19.285^{+0.025}_{-0.027}$ \\[6pt]
\hline\hline
\end{tabular}
\end{table}
% \clearpage
 it can be observed that the CC data provide an estimate of $H_0 = 65.438 \pm 0.088$ km/s/Mpc, which is more consistent with the Planck 2018 CMB result of $67.4 \pm 0.5$ km/s/Mpc. On the other hand, the Pantheon+SH0ES dataset gives a higher value of $H_0 = 73.0^{+1.0}_{-1.1}$ km/s/Mpc, in agreement with local observations. It is noteworthy that the framework accommodates both estimates simply by changing the dataset. This demonstrates its robustness and independence from specific assumptions, while the variation itself highlights the ongoing Hubble tension in cosmology.

\begin{figure}[h!]
\centering
		\includegraphics[height=7.5cm,width=9.0cm]{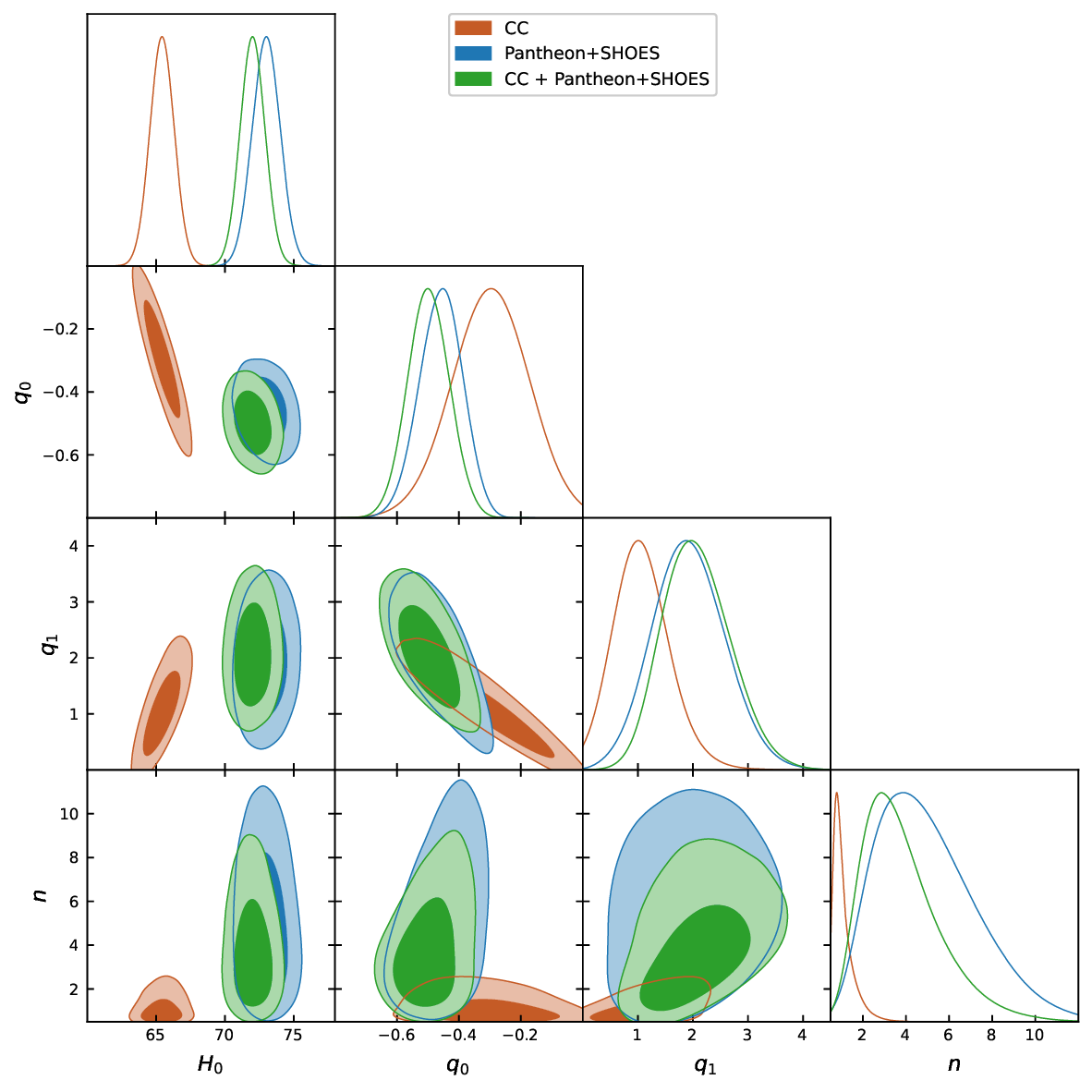}
		\caption {The posterior distribution of various observational data measurements using the HL model, indicating the regions within $1\sigma$ and $2\sigma$. } 
\label{f6}
\end{figure}
\clearpage
\section{Dynamical Behavior of Cosmological Parameter across Redshift}\label{s4}
We now focus on analyzing the intrinsic properties of the parameters through Eqs. (\ref{7}) and (\ref{8}). By substituting the expression for $H(z)$, we arrive at the following forms of energy density and pressure:
\begin{equation}\label{17}
    \rho = \frac{6(3\lambda - 1)}{\kappa^2} H_0^2 (1+z)^{2+2q_0} 
\left[\frac{(1+z)^n}{1 + n\ln(1+z)}\right]^{2q_1/n^2} 
- \frac{3\kappa^2 \mu^2 \Lambda^2}{8(3\lambda - 1)} 
- \frac{3\kappa^2 \mu^2 k^2 (1+z)^4}{8(3\lambda - 1)} 
+ \frac{3\kappa^2 \Lambda \mu^2 k (1+z)^2}{4(3\lambda - 1)}
\end{equation}
\begin{equation}\label{18}
\begin{split}
p = {} & -\frac{6(3\lambda - 1)}{\kappa^2} H_0^2 (1+z)^{2(1+q_0)} 
\left[\frac{(1+z)^n}{1 + n\ln(1+z)}\right]^{2q_1/n^2} \\
& + \frac{4(3\lambda-1)}{\kappa^2} H_0^2 (1+z)^{2(1+q_1)} 
\left[\frac{(1+z)^n}{1 + n\ln(1+z)}\right]^{2q_1/n^2} \\
& - \frac{3\kappa^2 \mu^2 \Lambda^2}{8(3\lambda - 1)} 
- \frac{3\kappa^2 \mu^2 k^2 (1+z)^4}{8(3\lambda - 1)} 
- \frac{\kappa^2 \Lambda \mu^2 k (1+z)^2}{2(3\lambda - 1)}
\end{split}
\end{equation}

\begin{figure}[h!]
\centering
\begin{subfigure}[b]{0.48\textwidth}
    \centering
    \includegraphics[height=6.0cm,width=\textwidth]{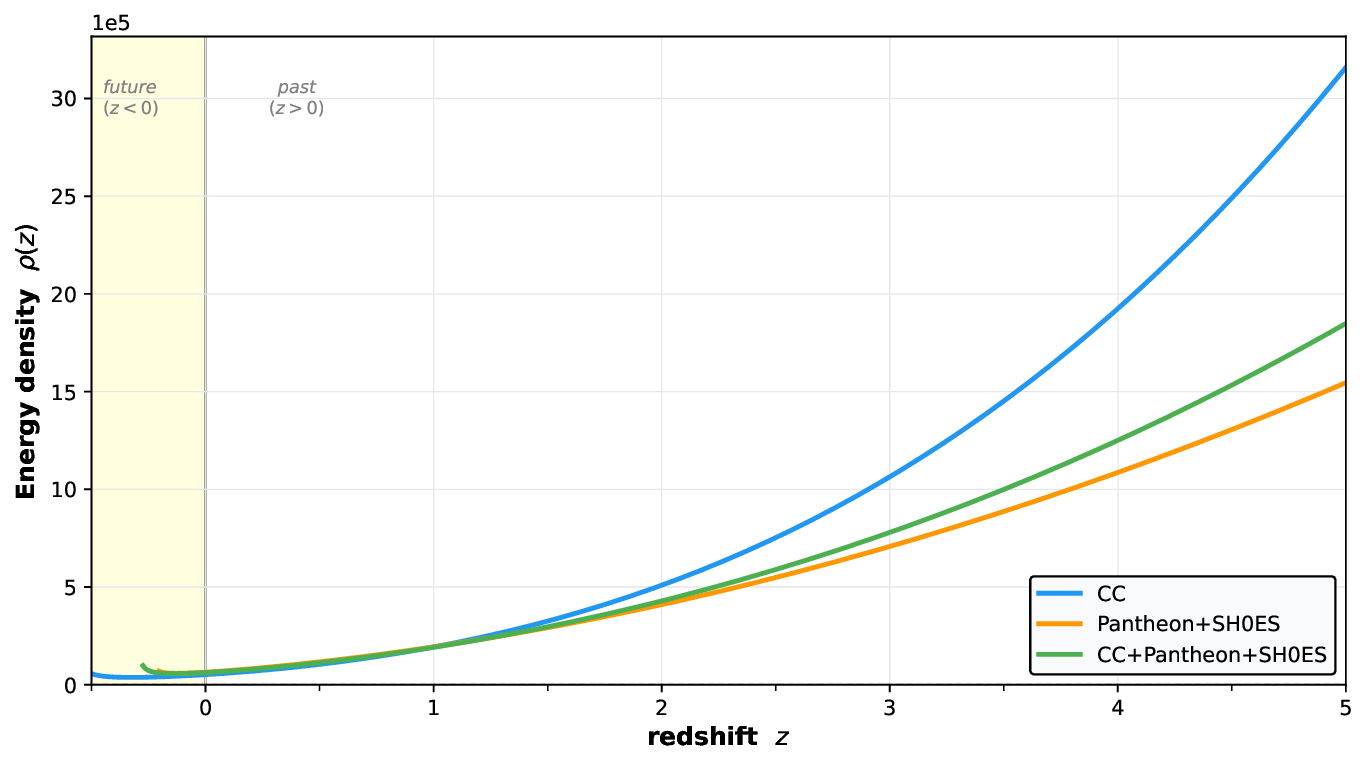}
    \caption{$\rho$ vs $z$}
    \label{f7a}
\end{subfigure}
\hfill
\begin{subfigure}[b]{0.48\textwidth}
    \centering
    \includegraphics[height=6.0cm,width=\textwidth]{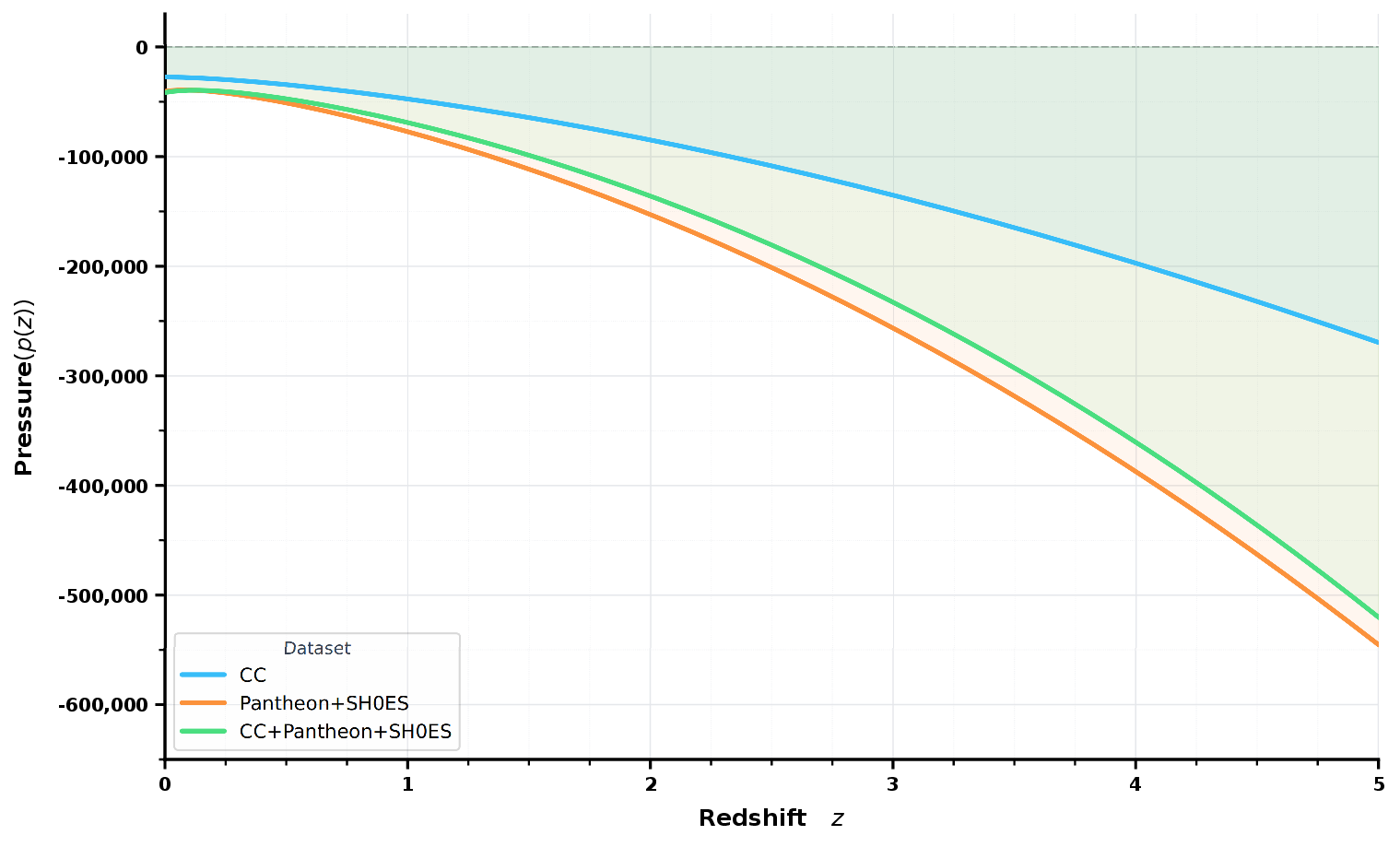}
    \caption{$p$ vs $z$}
    \label{f7b}
\end{subfigure}
\caption{Evolution of energy density and pressure with redshift $z$.}
\label{f7}
\end{figure}
The evolution of energy density and pressure with redshift z for CC, Pantheon+SH0ES, and their combination is presented in Figure \ref{f7}. As seen in Figure \ref{f7a}, the energy density $\rho(z)$ attains large positive values at high redshift, reflecting the dense state of the early universe. With cosmic expansion toward lower redshift, it decreases smoothly in a monotonic manner while remaining strictly positive throughout. As $z \to -1$, the energy density gradually approaches zero, indicating a continuously expanding universe evolving toward a de Sitter-like state. Similarly, the variation of pressure $p(z)$ with redshift for the three datasets is presented in Figure \ref{f7b}. The negative nature of pressure persists throughout the evolution and intensifies at lower redshifts, consistent with dark energy–driven expansion. At the present epoch $(z=0)$, the high negative pressure indicates the dominance of repulsive gravitational effects. The close agreement between the three datasets further confirms that the proposed parametrization performs reliably within the  Ho\v{r}ava-Lifshitz setting.

\begin{figure}[htbp]
\centering

\begin{subfigure}[b]{0.50\textwidth}
    \centering
    \includegraphics[width=\linewidth]{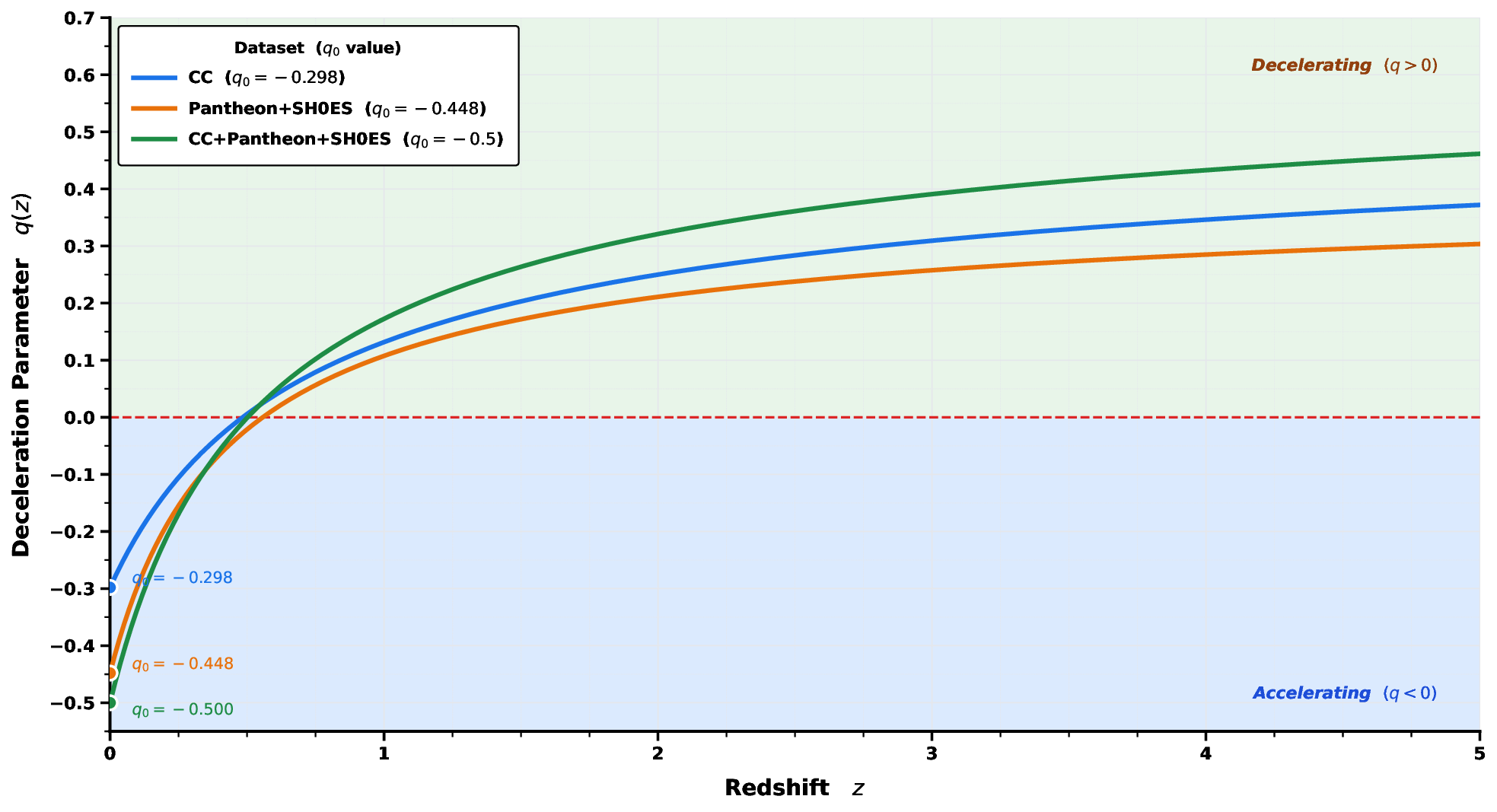}
    \caption{$q(z)$ vs $z$}
    \label{f8a}
\end{subfigure}\hfill
\begin{subfigure}[b]{0.47\textwidth}
    \centering
    \includegraphics[width=\linewidth]{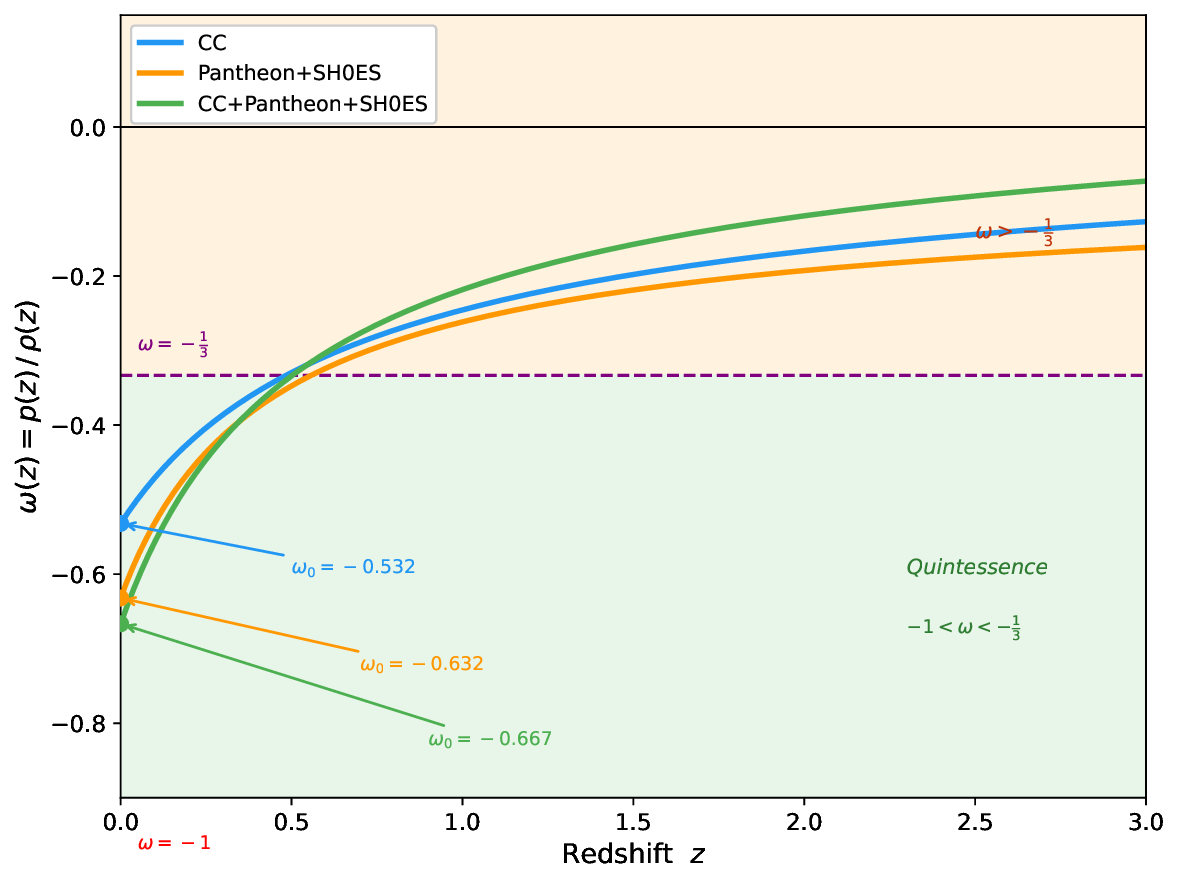}
    \caption{$w(z)$ vs $z$}
    \label{f8b}
\end{subfigure}

\caption{Evolution of cosmological parameters with redshift $z$.}
\label{f8}

\end{figure}
The evolution of the deceleration parameter q(z) as a function of redshift is shown in Figure \ref{f8a}. The behaviour of the model is in good agreement with the standard $\Lambda$CDM scenario throughout the considered redshift range. The positive values of the deceleration parameter at high redshift reflect a decelerating universe dominated by matter. With cosmic evolution, the deceleration parameter shifts gradually from positive to negative values, marking the transition from deceleration to acceleration driven by dark energy. At the present epoch, the values of the deceleration parameter are $q_0 = \text{-0.298}$ for CC, $q_0 = \text{-0.448}$ for Pantheon+ SHOES, and $q_0 = \text{-0.5}$ for their joint analysis. The negative values of the deceleration parameter confirm that the universe is currently in an accelerating phase. For the CC, Pantheon+SHOES and combined datasets, a clear transition from positive to negative values is observed, highlighting the change from decelerated to accelerated expansion. 
\par The variation of the  equation of state parameter (EoS) $\omega= p/\rho$  with redshift is given in Figure \ref{f8b}. The present-day values are $\omega_0 \approx -0.532$ for CC, $\omega_0 \approx -0.632$ for Pantheon+SH0ES, and $\omega_0 \approx -0.667$ for their joint analysis. A consistent behaviour is seen across all datasets: the EoS parameter starts with less negative values at earlier times and shifts to more negative values today, reflecting the growing influence of dark energy. For the full redshift range, $\omega$ stays in the quintessence interval $(-1 < \omega < -1/3)$, confirming that the observed acceleration is caused by a dynamical dark energy component rather than a cosmological constant. The functional form in Eq.(\ref{19}) suggests that $\omega$ may approach $-1$ in the asymptotic future $(z \to -1)$, indicating a de Sitter-like state, even though this region lies beyond the scope of the present analysis. It is important to note that None of the datasets predict $\omega < -1$, indicating that phantom dark energy is not supported in this framework.
{\small
\begin{equation}\label{19}
\omega(z) = \frac{
-\frac{6(3\lambda-1)}{\kappa^2}H_0^2(1+z)^{2(1+q_0)}\mathcal{F}(z)
+\frac{4(3\lambda-1)}{\kappa^2}H_0^2(1+z)^{2(1+q_1)}\mathcal{F}(z)
-\frac{3\kappa^2\mu^2\Lambda^2}{8(3\lambda-1)}
-\frac{3\kappa^2\mu^2 k^2(1+z)^4}{8(3\lambda-1)}
-\frac{\kappa^2\Lambda\mu^2 k(1+z)^2}{2(3\lambda-1)}
}{
\frac{6(3\lambda-1)}{\kappa^2}H_0^2(1+z)^{2+2q_0}\mathcal{F}(z)
-\frac{3\kappa^2\mu^2\Lambda^2}{8(3\lambda-1)}
-\frac{3\kappa^2\mu^2 k^2(1+z)^4}{8(3\lambda-1)}
+\frac{3\kappa^2\Lambda\mu^2 k(1+z)^2}{4(3\lambda-1)}
}.
\end{equation}
}
where,
\begin{equation*}
\mathcal{F}(z) = \left[\frac{(1+z)^n}{1+n\ln(1+z)}\right]^{2q_1/n^2},
\end{equation*}

\begin{figure}[h!]
\centering
		\includegraphics[height=7.0cm,width=11.0cm]{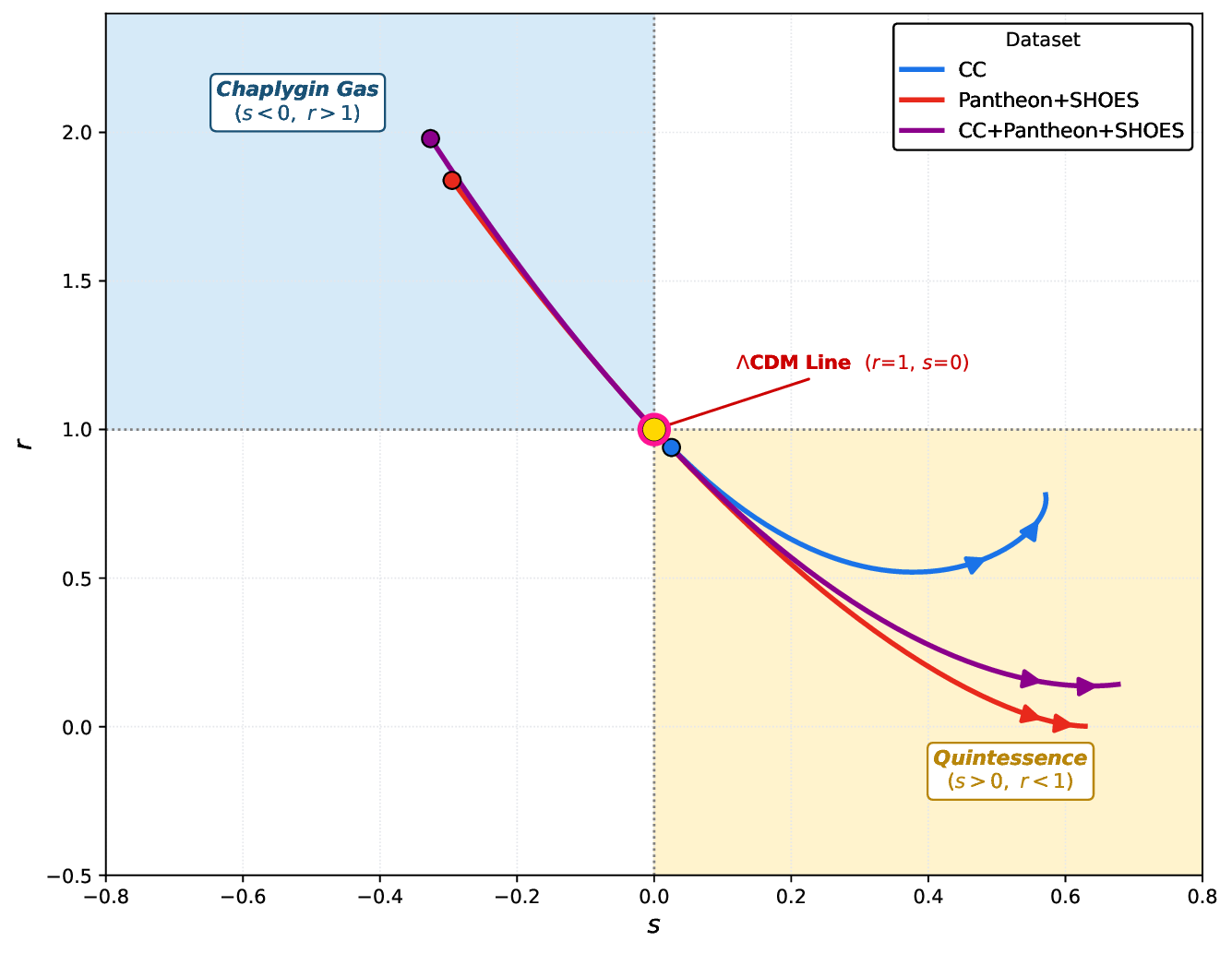}
		\caption {Representation of the ${r-s}$ trajectories for the HL model. } 
\label{f9}
\end{figure}\
In cosmology, understanding the evolution of the Universe requires a careful study of dark energy and its role in cosmic expansion. To examine this without assuming a specific dark energy model, cosmologists use a useful tool known as the statefinder diagnostic and is denoted by $\{r, s\}$. As proposed in \cite{sahni2003,bamba2012,alam2003}, this mathematical framework relies on higher derivatives of the cosmic scale factor to characterize expansion and provides a useful way to compare different dark energy scenarios. According to Sahni et al. \cite{sahni2003}, these parameters can be used to categorize cosmological models into quintessence and Chaplygin gas classes. The $\Lambda$CDM model is characterized by the fixed point $\{r, s\} = \{1, 0\}$. In contrast, quintessence models are located in the region $r < 1,\, s > 0$, while Chaplygin gas models are found in the region $r > 1,\, s < 0$ of the $r-s$ plane. For the present model the $\{r, s\}$ parameters are given by
\begin{equation}\label{20}
    r = 2q^2 + q + (1+z)\frac{dq}{dz}\,
\qquad
s = \frac{r - 1}{3\left(q - \dfrac{1}{2}\right)}\,
\end{equation}
from Equations. (\ref{11})and (\ref{20})
\begin{equation}\label{21}
    r(z) = 2\left[q_0 + \frac{q_1 \ln(1+z)}{1 + n\ln(1+z)}\right]^2 
     + \left[q_0 + \frac{q_1 \ln(1+z)}{1 + n\ln(1+z)}\right] 
     + \frac{q_1}{\left[1 + n\ln(1+z)\right]^2}
\end{equation}
\begin{equation}\label{22}
    s(z) = \frac{2q^2(z) + q(z) + \dfrac{q_1}{\left[1 + n\ln(1+z)\right]^2} - 1}
       {3\left(q_0 + \dfrac{q_1\ln(1+z)}{1 + n\ln(1+z)} - \dfrac{1}{2}\right)}
\end{equation}

The present model is analyzed using the statefinder parameters ${r, s}$, and the resulting trajectories are presented in Figure \ref{f9}. The trajectories for CC, Pantheon+SH0ES, and the combined dataset initially lie in the Chaplygin gas region, approach the $\Lambda$CDM point, and eventually move into the quintessence region at late times. This evolution through the $\Lambda$CDM point demonstrates that the model follows standard cosmological behaviour at intermediate times, while its shift into the quintessence region at late times is in agreement with the EoS results in Figure \ref{f8b}. The fact that all dataset trajectories evolve along a common path indicates that the proposed parametrization is robust within the Ho\v{r}ava-Lifshitz framework.

\begin{figure}[h!]
\centering
		\includegraphics[height=7.0cm,width=11.0cm]{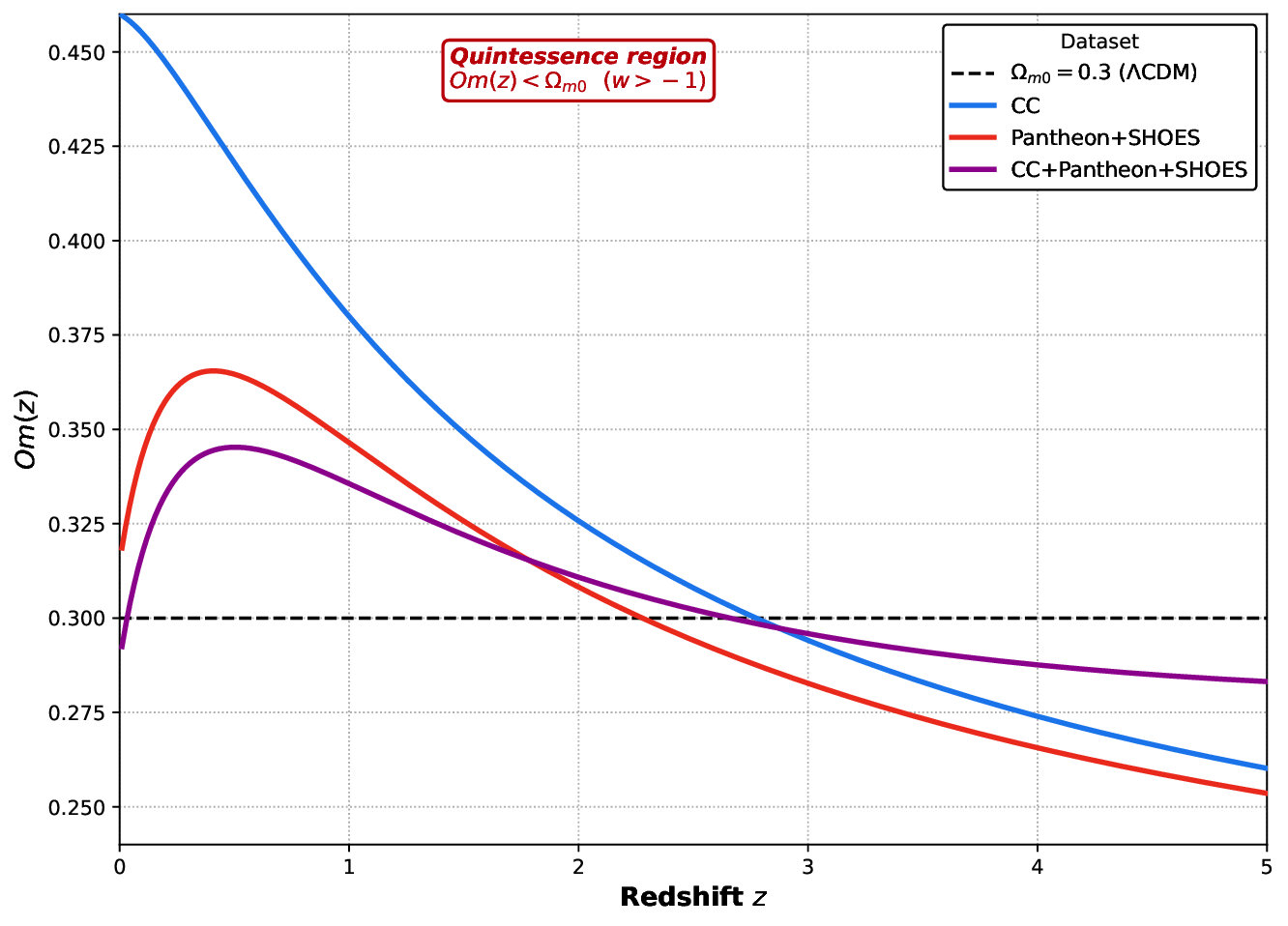}
		\caption {Behaviour of the $Om(z)$ diagnostic as a function of $z$ for the HL model.  } 
\label{f10}
\end{figure}
The $Om(z)$ diagnostic is described in Figure \ref{f10} as a useful model-independent probe of dark energy. It depends only on the Hubble parameter $H(z)$ and is defined as
\begin{equation}\label{23}
Om(z) = \frac{\left[\frac{H(z)}{H_0}\right]^2 - 1}{(1+z)^3 - 1}.
\end{equation}
In the $\Lambda$CDM scenario, $Om(z)$ does not change with redshift and remains fixed at $\Omega_{m0}$, marked by the dashed horizontal line at $\Omega_{m0} = 0.3$. The sign of the slope of $Om(z)$ distinguishes different dark energy behaviours: a negative slope indicates quintessence-like ($w > -1$), while a positive slope points to phantom-like energy($w < -1$). The figure \ref{f10} clearly shows that $Om(z)$ decreases with redshift for all three datasets within $z \in [0, 5]$, supporting a quintessence-type dark energy scenario. The CC dataset starts at a higher initial value $(Om(0) \approx 0.46)$ and decreases continuously, crossing the $\Omega_{m0} = 0.3$ reference around $z \approx 3$. On the other hand, the Pantheon+SH0ES and combined datasets stay close to the $\Lambda$CDM line, showing a mild increase at low $z$ before turning downward and eventually dropping below $\Omega_{m0} = 0.3$. The clear deviation of $Om(z)$ from the $\Lambda$CDM constant value, combined with its persistent negative slope at higher redshifts, rules out a cosmological constant interpretation and supports a quintessence-type description of late-time cosmic acceleration. 
\begin{figure}[h!]
\centering
		\includegraphics[height=7.0cm,width=11.0cm]{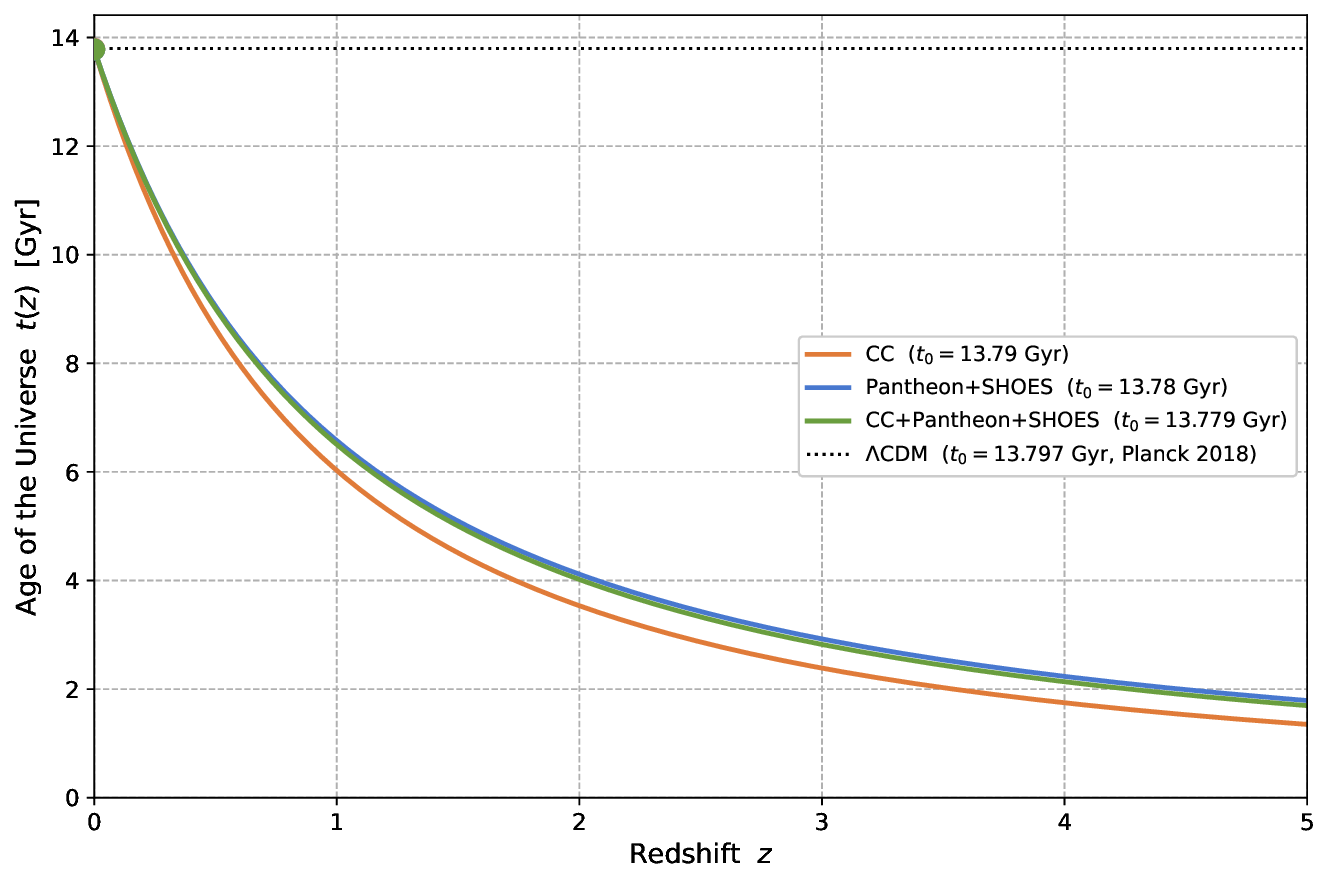}
		\caption {Age evolution of the universe for the CC, CC+Pantheon+SHOES, and CC+Pantheon+SHOES datasets as a function of redshift $z$. } 
\label{f11}
\end{figure}
 \par To determine the present-day age of the Universe $t_0$, we evaluate the integral given below.
\begin{equation}\label{24}
    t_0 = \int_0^{\infty} \frac{dz}{(1+z)\,H(z)}
\end{equation}
By employing the constrained parameters for each dataset, the resulting age–redshift relations are presented in Figure \ref{f11}. The estimated cosmic age at the present epoch is $t_0 = 13.79$ Gyr (CC), $13.78$ Gyr (Pantheon+SH0ES), and $13.779$ Gyr for the joint dataset. There is strong agreement between these results and the Planck 2018 $\Lambda$CDM estimate of $t_0 = 13.797$ Gyr, shown by the dotted line. The strong consistency between the estimated ages from all datasets and the Planck result further validates the logarithmic $q(z)$ parametrization as a viable description of the late-time Universe in Ho\v{r}ava–Lifshitz gravity.

\section{Conclusion}\label{s5}
We examine the late-time dynamics of the Universe within Ho\v{r}ava–Lifshitz gravity by adopting a model-independent kinematic approach. To keep the analysis model-independent, we introduce a logarithmic form of the deceleration parameter, $  q(z) = q_0 + \frac{q_1 \ln(1+z)}{1 + n \ln(1+z)}$, which is finite for all redshifts and effectively describes the transition from a decelerating Universe in the past to an accelerating one at present. Using the ansatz in Eq. (\ref{12}), we derived $H(z)$ analytically and constrained the parameters $H_0$, $q_0$, $q_1$, and $n$ through MCMC analysis to CC, Pantheon+SH0ES, and combined datasets. The resulting best-fit values in Table \ref{t1} consistently give negative $q_0$, indicating that the Universe is currently undergoing accelerated expansion. At low redshifts, the reconstructed $H(z)$ agrees well with $\Lambda$CDM, with only mild differences emerging at higher redshifts (Figures \ref{f1} and \ref{f2}). The posterior plots (Figures \ref{f3}-\ref{f6}) further confirm that the parameters are tightly constrained and consistent among all datasets.
\par The results of the dynamical analysis are in line with physical expectations. The energy density $\rho(z)$ remains positive and decreases monotonically, whereas the pressure $p(z)$ stays negative and strengthens at lower redshifts, both supporting a dark energy–driven expansion scenario (Figure \ref{f7}). The behaviour of $q(z)$ indicates a smooth transition from deceleration at high redshifts to acceleration at present (Figure \ref{f8a}), with the transition redshift falling within the expected range $z_t \approx 0.6$-$0.8$. Moreover, a smooth evolution is observed in $\omega(z)$, moving from less negative values at earlier times to more negative values at the present epoch, yielding $\omega_0 = -0.532$ for CC, $-0.632$ for Pantheon+SH0ES, and $-0.667$ for the combined dataset. Furthermore, it can be easily observed that at all redshifts, $\omega$ stays within the quintessence domain ($-1 < \omega < -1/3$), suggesting that the model does not favor phantom dark energy (Figure \ref{f8b}).
\par Figure \ref{f9} presents the statefinder $\{r, s\}$ trajectories, which are consistent with our earlier findings. All three datasets start in the Chaplygin gas region, approach the $\Lambda$CDM fixed point, and then evolve into the quintessence region at late times. The crossing near the $\Lambda$CDM point confirms that the model recovers standard cosmology during intermediate epochs. As seen in Figure \ref{f10}, the $Om(z)$ diagnostic exhibits a persistent negative slope for all datasets, consistent with quintessence-like behaviour. The CC curve decreases monotonically from $Om(0) \approx 0.46$, while the Pantheon+SH0ES and combined datasets remain near $\Omega_{m0} = 0.3$, showing a small rise at low redshifts before declining. The non-constant nature of $Om(z)$ clearly rules out the cosmological constant model.
\par As a final remark, the estimated age of the Universe at present is $t_0 = 13.79$ Gyr for CC, $13.78$ Gyr for Pantheon+SH0ES, and $13.779$ Gyr for the joint dataset, showing strong consistency with the Planck 2018 $\Lambda$CDM estimate of $13.797$ Gyr and can be seen in Figure \ref{f11}.
\par In conclusion, the logarithmic $q(z)$ parametrization provides a simple yet effective way to describe the transition from early deceleration to present acceleration within Ho\v{r}ava–Lifshitz gravity, without the need for specific dark energy assumptions or additional free parameters. The agreement among various observational probes and diagnostics- including the EoS, statefinder, and $Om(z)$ - provides strong evidence that this framework can serve as an effective alternative to $\Lambda$CDM in describing the Universe’s late-time behavior.

 \section*{\textbf{Acknowledgment}}
  Shivani, acknowledges CSIR-UGC, New Delhi, for the financial aid provided under the CSIR-UGC(JRF) scheme with the UGC-Ref.No.: 1332/(CSIR-UGC NET JUNE 2019). RC thanks SERB, New Delhi, for financial assistance through project No. CRG/2023/004560 (P-07/1328). 
\section*{Data Availability}

There are no new data associated with this article.

\small
\bibliographystyle{ieeetr}
\bibliography{referenceHL4}
\end{document}